\documentclass[aps,prl,twocolumn,groupedaddress]{revtex4}

\newcommand{\ket}[1]{\left| #1 \right>} 
\newcommand{\bra}[1]{\left< #1 \right|} 
\newcommand{\bracket}[2]{\left< #1 \;\middle|\; #2 \right>}

\usepackage{mathrsfs}  
\usepackage{amssymb,latexsym,amsmath}
\usepackage{graphicx}
\usepackage{ulem}

\begin{document}

\preprint{LA-UR-xxxxx}

\title{Entangled-photon interferometry for plasmas}


\author{Zhehui Wang}
\email[]{zwang@lanl.gov}
\affiliation{Los Alamos National Laboratory, Los Alamos, NM 87545, USA}

\author{Yanhua Shih}
\affiliation{University of Maryland, Baltimore County, Baltimore, MD 21250, USA}


\date{\today}

\begin{abstract}
Sub-picosecond coincidence timing from nonlocal intensity interference of entangled photons allows quantum interferometry for plasmas. Using a warm plasma dispersion relation, we correlate phase measurement sensitivity with different plasma properties or physics mechanisms over 6 orders of magnitude. Due to $N^\alpha$ ($\alpha \leq -1/2$) scaling with the photon number $N$, quantum interferometry using entangled light can probe small signals in plasmas not previously accessible. As an example, it is predicted that plasmas will induce shifts to a Gaussian dip, a well-known quantum optics phenomenon that is yet to be demonstrated for plasmas.
\end{abstract}



\maketitle

LIGO's interferometer has reached a phase sensitivity equivalent to 10$^{-5} \times$ the width of a proton~\cite{AAA:2013}. Even though such a sensitivity may not be required in many plasma experiments, we show here that by reaching a phase sensitivity down to $\sim 1\times $ proton width, a variety of plasma properties or physics mechanisms not accessible until now may become directly observables. Observation of small signals is of broad interest, not only to direct comparisons between experiments with theory and simulations, but also to predictive plasma science and technology, including, for example, harvesting high-temperature plasmas as a carbon-free and sustainable energy source.  Small amplitude fluctuations in plasma density and magnetic fields are universal signatures for enhanced transport, particle and energy loss in turbulent plasmas~\cite{Ped:1999,BDT:2003, MM:2011}. Identification of new precursors based on even smaller signals will have significant implications to prediction of disruptions in tokamak plasmas for fusion energy, to inertial confinement plasmas for initiation and growth of Rayleigh-Taylor instability and Richtmyer-Meshkov instability, to plasma processing of large-area (300-mm diameter) silicon wafers with feature sizes below 10 nm, and to laboratory astrophysics experiments with high Lundquist number ($> 10^7$) or large magnetic Reynolds number ($>10^4$). Predictions of plasma dynamics remain to be a grand challenge even through modern computing and growingly sophisticated measurement tools~\cite{HP:2005,Filbet:2006,Evans:1976,Kunze:2005,DCB:2007,DKK:2014}. Most instruments and measurements, operating on classical physics principles, are far from theoretically allowed limits in sensitivity.

We introduce quantum interferometry that uses entangled photons as sensitive probes for small amplitude fluctuations in plasmas, or `an interferometer for turbulence'.  Optical interferometers are widely used in experimental science including plasma experiments~\cite{MW:1995,SZ:1997,Glauber:2007,Shih:2011,Hut:2002}. Prior to gravitational wave detection, Michelson interferometer, Mach-Zehnder interferometer, and Hanbury-Brown-Twiss stellar interferometer have played pivotal roles in the 20th century experimental physics~\cite{HanB:1974}. Recent advances in entangled light sources and quantum optics have stimulated the growth of quantum interferometry for experimental physics~\cite{Shih:2003,PCL:2012}.  Several novel features of the entangled-photon interferometers are noted: (a) the light source is based on entangled photons rather than lasers; (b) phase detection is based on nonlocal correlation in intensity rather than correlation in field or amplitude; and (c) phase detection sensitivity is scalable with the number of photons according to the quantum limits, $\propto N^\alpha$ ($\alpha \leq -1/2$) for the photon number $N$. We use 1 $\mu$m photon as a primary example below for quantitative estimates. This does not preclude the use of the photons in the other parts of the electromagnetic spectrum for entangled photon interferometry, as long as the three conditions can be met.

{\it Phase modulation in a plasma}
Because of the weak interactions between plasma electrons, ions, atoms, and atomic agglomerates  with the two-photon state from an entangled photon source, we may adopt the approximation that a plasma only modulates but does not destroy quantum information such as photon entanglement. A plasma may induce a phase change $\Phi_P$ to photon wavefunctions. A pure real number $\Phi_P$ corresponds to a unitary transformation of entangled photon wavefunctions that preserves the probability amplitude. 

We first show that, through its dependence on plasma refractive index $n$, $\Phi_P = \omega \int n dl/c$, $\Phi_P$ can vary with a wide set of plasma parameters. Here $\int dl$ is a path through a plasma, $\omega$ the photon angular frequency, and $c$ the speed of light in vacuum. $\Phi_P = \Phi_P (n) = \Phi_P(n_e, n_i, {\bf B}, {\bf v}_e, {\bf v}_i, T_e, T_i, \cdots)$, where $n_e$,  $n_i$, {\bf B}, $T_e$, $T_i$, {\bf v}$_e$, and {\bf v}$_i$ symbolize electron density, ion density, magnetic field, electron flow, ion flow, respectively. Quantum measurements of $\Phi_P$, and especially its derivatives such as $\partial \Phi_P/ \partial n_e$, $\partial \Phi_P/\partial B_i$, $\partial \Phi_P/ \partial T_e$, {\it etc}. are the basis of quantum metrology for plasmas. The rest of the paper will focus on entangled-photon interferometry to measure $\Phi_P$ and its variations. 

For simplicity, we assume that the plasma frequency $\omega_{pe}$ is well below the cutoff frequency for the entangle photons, $\omega_{pe} = \sqrt{\frac{e^2 n_e}{\epsilon_0 m_e}} << \omega$, so that entangled photons pass through a plasma without being reflected. Here $\epsilon_0$ is the vacuum permittivity without the plasma, $e$ the elementary charge, and $m_e$ the electron rest mass. We may treat plasma effects on the entangled photons classically according to Maxwell's equations, and obtain in terms of electric field {\bf E},
\begin{equation}
-\nabla^2 {\bf E} + \nabla (\nabla \cdot {\bf E}) -\epsilon_0 \frac{\omega^2}{c^2} {\bf E} = \frac{\omega^2}{c^2} \boldsymbol{\epsilon}_{P} \cdot {\bf E}. 
\label{eq:maxwell}
\end{equation}
 $\boldsymbol{\epsilon}_{P}$ is in general a $3\times 3$ matrix for the anisotropic plasma dielectric coefficients.  Each matrix element may be a sum of  contributions from free electrons and ions. $\boldsymbol{\epsilon}_{P}$ may also include effects from charge-neutral materials such as a dielectric solid, liquid, or gas mixed with a plasma. 

For electromagnetic wave propagation parallel to a uniform static magnetic field ${\bf B}_0$, the plasma refractive index is dispersive for different frequencies and wavelengths  and has been derived by solving the Vlasov equations for  $\boldsymbol{\epsilon}_{P}$~\cite{KP:1966,Sti:1992},
\begin{equation}
\begin{split}
n_{R,L} =& 1 + \sum_s \frac{\omega_{ps}^2}{\omega^2} \int_{-\infty}^\infty dv_\| \int_0^\infty \pi v_\perp^2 dv_\perp \frac{1}{\omega - k_\|v_\| \pm \Omega_s} \\
& \times \left[(\omega-k_\|v_\|) \frac{\partial F_s}{\partial v_\perp} + k_\| v_\perp \frac{\partial F_s}{\partial v_\|}\right].
\end{split}
\end{equation}
A magnetized plasma induces birefringence for light propagation depending on the polarization of the light. $n_{R,L} = kc/\omega$ are the refractive indices for right (R) and left (L) hand circular polarizations respectively. Here $k = k_\|$ (the wave vector component parallel to the magnetic field) or $k_\perp = 0$ (the wave vector perpendicular to the magnetic field) for the photons propagating along the magnetic field. The plus/minus sign corresponds to the right/left-hand circular polarization of light. $F_s$ is the normalized distribution for $s$-species (electrons and different types of ions), $f_s = n_s F_s(v_\|, v_\perp)$, and $\int d{\bf v} F_s$ = 1. The {\bf k} vector aligns with {\bf B}$_0$, which is assumed to be in the $z$-direction. The refractive indices $n_{R, L} $ can be further reduced to be a function of the plasma dispersion function $Z_0(\zeta_\pm)$~\cite{Sti:1992}. Since $\lambda = \frac{1}{2} k_\perp^2 \rho_L^2 = 0$ for $k_\perp = 0$, finite Larmor radius effects do not appear, $\rho_L$ is the Larmor radius for an electron or an ion, even though $\frac{1}{2} k_\|^2 \rho_L^2 \ge 1$ for visible and near infrared light.

Since for an electron or an ion, $\zeta_\pm = \frac{\omega-k_\| v_\| \pm \Omega}{k_\| w_\|} \sim \frac{\omega}{k_\|w_\|}\gg 1$, or the plasma thermal and non-thermal motion are non-relativistic, the asymptotic expansion for the plasma dispersion function $Z_0(\zeta) \sim -\zeta^{-1} - \zeta^{-3} + \cdots$ is used in the Taylor expansion for large $\zeta_\pm$, $Z_0(\zeta)$ is slightly different from the Fried-Conte plasma dispersion function $Z(\zeta) \sim -\zeta^{-1}-\zeta^{-3}/2+\cdots$. The imaginary term ($\propto i e^{-\zeta^2}$) may be neglected for both $Z(\zeta)$ and $Z_0 (\zeta)$. 
Keep terms to the order $\omega^{-4}$, we obtain
\begin{widetext}
\begin{equation}
n_{R, L} =  1 -  \sum_s \frac{\omega_{ps}^2}{2\omega^2} \left[1 \mp \frac{\Omega_s}{\omega} + (\frac{\Omega_s^2}{\omega^2} \mp \frac{\Omega_s k_\|v_{\|s}}{\omega^2}+\frac{T_{s\perp} k^2_\|}{m_s\omega^2}) + O(\frac{1}{\omega^3}) \right],
\label{eq:disp}
\end{equation}
\end{widetext}
where `$-/+$' is for R/L-mode. The summation is over different charged particle species (electron and one or more ions). For the particle species $s$, $\omega_{ps}$, $\Omega_s$, $v_{\|s}$, $T_{s\perp}$  are the plasma frequency, gyrofrequency, flow along the magnetic field, and temperature perpendicular to the magnetic field {\bf B}$_0$, respectively. In a fusion plasma, for example, there could be deuterons, tritons (for DT fusion) and helium ions due to fusion. To the order $\omega^{-3}$, we recover the cold plasma approximation from Eq.~(\ref{eq:disp}). The flow and thermal effects do not show up until to the order $\omega^{-4}$, which is usually neglected for existing instruments~\cite{Hut:2002}. The electron plasma frequency contribution $ n(\omega_{pe})= \omega_{pe}^2/(2\omega^2) = e^2 n_e /(2\epsilon_0 m_e \omega^2) = 2\pi r_e c^2 n_e/\omega^2$, dominates over other terms because the electron mass is much less than the ions, corresponding to $n$($\omega_{pe}$) = 4.5$\times$ 10$^{-8}$ for $n_e = 10^{20}$ m$^{-3}$ and a probe photon wavelength of 1 $\mu$m (cutoff plasma density 1.1$\times 10^{27}$ m$^{-3}$). The photon transit time lag through a 1-m-long plasma column relative to vacuum would be $\tau (\omega_{pe}) = 0.15 $ fs, which is within the resolution of existing table-top entangled photon interferometers~\cite{LKB:2018}. For a laser-produced high-density plasma with an electron density in the range of 10$^{24}$-10$^{26}$ m$^{-3}$ and a plasma column length of 1 to 0.1 cm, the line-integrated density can reach 10$^{22}$ to 10$^{23}$ m$^{-2}$. Correspondingly, a photon travel time lag $\tau (\omega_{pe})$ is in the range of 15 to 150 fs, which has been demonstrated by a growing number of groups since the 1980s~\cite{HOM:1987,SA:1988,SSR:1994b,DJM:1999}.

We may similarly estimate the magnitude of smaller amplitude effects in Eq.~(\ref{eq:disp}) for 1-$\mu$m photons. For a hydrogen plasma, $n(\omega_{pH}) = \omega_{pH}^2/(2\omega^2) = \sqrt{m_e/m_H} n(\omega_{pe})=2.3\times 10^{-2} n(\omega_{pe})$. $m_H =$ 1836 $m_e$ is the proton mass. For an helium-4 ion ($Z=2$) density fraction of $f_\alpha$ =3\%, $n(\omega_{pHe}) = \omega_{pHe}^2/(2\omega^2) = Z\sqrt{f_\alpha m_e/m_{He}} n(\omega_{pe}) = 4.0 \times 10^{-3} n(\omega_{pe})$. $n(\Omega_e) = (\Omega_e/\omega) n(\omega_{pe}) = 9.3 \times 10^{-5} n(\omega_{pe})$ for  $B_0$ = 1 Tesla. $n(v_{\|e}) = (\Omega_s k_\|v_{\|s}/\omega^2) n(\omega_{pe}) = 1.4 \times 10^{-2} n(\Omega_{e})  = 1.3 \times 10^{-6} n(\omega_{pe}) $ for electron flow with a kinetic energy of 100 eV. $n(T_{\perp e}) = (T_{e\perp} k^2_\|)/(m_s\omega^2) n(\omega_{pe})$ = 2.0$\times 10^{-4} n(\omega_{pe})$. 

In short, increasing phase measurement sensitivity down to $10^{-6} n(\omega_{pe}) \sim 10^{-14}$, equivalent to a length sensitivity of 10$^{-14}$ m or comparable to 10$\times$ the width of a proton, would allow sensitive probing a broad range of plasma parameters and their variations. Here we assume 1 $\mu$m probe photon wavelength for 10$^{20}$ m$^{-3}$ plasmas. Using longer wavelength photons may improve the phase sensitivity due to $1/\omega^k$-dependence ($k \ge$ 2) in Eq.~(\ref{eq:disp}).

\begin{figure}[htbp] 
   \centering
   \includegraphics[width=3in]{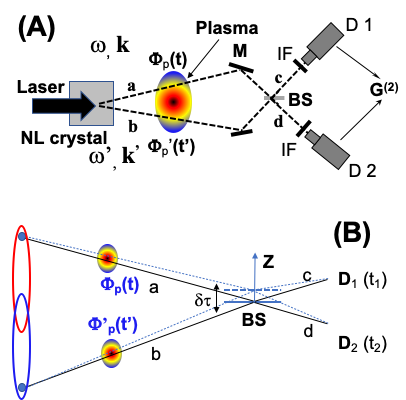} 
   \caption{(A) Schematic configuration of a single-chord biphoton interferometery for plasma phase meaurement $\Phi_p$ based on the $G^{(2)}$ correlation function. Entangled photon pairs ({\bf k},{\bf k}$'$) are generated by pumping a nonlinear crystal with a laser, BS1 -- : beam splitter, IF: interference filters, D 1 -- 2: Detectors with single-photon sensitivity. (B) Simplified diagram for (A) that illustrates the two path-configurations and adjustment of the beam splitter (not to scale for visibility) for intensity interference.}
   \label{fig:ESI}
\end{figure}

In phase measurement reaches the quantum limits~\cite{GLM:2011}, measurement sensitivity is given by
\begin{equation}
\Phi_P = \frac{\omega}{c} \int n dl = \omega \tau_p \geq k_0 (\eta N)^\alpha,
\end{equation}
where $N$ is the quanta or photon used, $\alpha = -1/2$ is the shot noise limit or standard quantum limit~\cite{CTD:1980}, $\alpha = -1$ is the Heisenberg limit. $k_0 = 1/2$ in both limits. $\eta$ is the single-photon detection efficiency. Heisenberg limit may be achieved for entangled photons~\cite{HB:1993}. The existing experimental data indicates that current setup (see Fig.~\ref{fig:ESI} below without a plasma)  is $k_0 \sim 400$ for $\tau_p = 10$ fs ($\eta N = 100$), which is about 10$^3$ above the theoretical limits. 

{\it Quantum metrology for plasmas using entangled photon pairs}
We first summarize the theoretical framework to measure plasma phase-shift $\Phi_P$ through second-order correlation in intensity detection, or intensity interferometry. As in quantum optics and quantum field theory~\cite{MW:1995,SZ:1997,Shih:2011}, measurement of a photon field may be described by the second order correlation function $G^{(2)}(1; 2)$ ~\cite{Glauber:2007}, 
\begin{equation}
G^{(2)}(1; 2)  = \langle {E}^{(-)}(1) {E}^{(-)}(2) {E}^{(+)}(2) {E}^{(+)}(1) \rangle,
\end{equation}
which is historically stimulated by the Hanbury-Brown-Twiss experiment. $\langle O \rangle \equiv Tr[\rho O]$ is the statistically averaged measurement of operator $O = {E}^{(-)}(1) {E}^{(-)}(2) {E}^{(+)}(2) {E}^{(+)}(1)$. Here and below, we may use abbreviated notations of `1' and `2' and suppress explicit dependence on the position and time of the two separate detector pairs D1 $(\mathbf{r}_1, t_1)$ and  D2 $(\mathbf{r}_2, t_2)$.
As in quantum optics, the photon detection is described by the electric field operators $ {E}^{(+)} (i) \equiv {E}^{(+)} ({\bf r}_i, t_i)$ (i=1, 2) and their adjoint
${E}^{(-)} (i) = \left[ {E}^{(+)} (i) \right]^\dagger$. $\ket{\Psi}$ represents the wavefunction of the entangled two-photon or bi-photon using Dirac ket/bra notation, with $\bra{\Psi}$ being the complex conjugate, satisfying the normalization condition $\bracket{\Psi}{\Psi}=1$. When we only consider a field $\ket{\Psi}$ of photon pairs, $G^{(2)}(1; 2) = \big{|} A(1; 2) \big{|}^2$, and the complex detection amplitude $A(1; 2)$ is
\begin{align}
A(1; 2) = \langle 0 | {E}^{(+)}(2) {E}^{(+)}(1) | \Psi \rangle.
\label{eq:a12}
\end{align}
$\ket{0}$ corresponds to the vacuum state of the photon field. It is clear that $A(1;2) = A(2;1)$ since $E^{+}(2)$ commutes with $E^{+}(1)$, and  $G^{(2)}(1; 2) = G^{(2)}(2; 1)$. 

We now evaluate $G^{(2)}(1;2)$ for the configuration shown in Fig.~\ref{fig:ESI}, which uses entangled photon pairs generated by spontaneous parametric down conversion (SPDC) of a pump laser in a nonlinear
crystal~\cite{Shih:2003}. The SPDC process is referred to as type-I, if the two arms of a two-photon state, called signal and idler photon respectively, has identical
polarizations, and type-II, if they have orthogonal polarizations. The process is said to be
degenerate if the SPDC photon pair have the same free space wavelength or wave number,
and nondegenerate otherwise. This configuration can be made insensitive to the photon polarization by selecting the beam splitter and IF filters. 

For entangled source function~\cite{RKS:1994}
 \begin{equation}\label{biphoton-01}
\ket{\Psi} = \ket{0} +  \sum_{\omega  \omega'} \delta(\omega + \omega' - \omega_{p}) F_{\omega \omega'} \hat{a}^{\dag}(\omega) \hat{a}^{\dag}(\omega') \ket{0},
\end{equation}  
the probability amplitude is obtained, after introducing a new variable $\nu$, so that $\omega = \omega_p/2 +\nu$ ($+$) and $\omega' = \omega_p/2 - \nu$ ($-$),
\begin{equation} \begin{split}
& A(t_1; t_2) =    e^{i(\omega_p /2) (t_1 +t_2 - 2\tau_0)} \times \\
& [ T  e^{-i \omega_p/2 (-\tau_p' (t_1)-\tau_p(t_2))} \\
& \times\sum_\nu e^{-i\nu (t_2-t_1-\tau_p(t_2) + \tau_p'(t_1)))} F_\nu f_1(-) f_2(+) \\
& -R e^{-i \omega_p/2 (-\tau_p' (t_2)-\tau_p(t_1))} \\
&\times \sum_\nu e^{-i\nu (t_1-t_2 + 2\delta \tau-\tau_p(t_1 +\delta \tau)+\tau_p'(t_2 -\delta \tau))} F_\nu f_1(+) f_2(-) ]. 
\end{split} \end{equation}
$\sqrt{T}$ and $\sqrt{R}$ are the transmission and reflection coefficients of the beamsplitter BS1.  $R+T = 1$ for a lossless beam splitter. $f_j (\omega)$ ($j$ =1 ,2) are the band pass filter function in front of the detector $D_j$.The plasma phase-shift $\Phi_P(\omega)$ (or equivalently time lag $\tau_P$) only appears in the path $a$. A phase compensation component (P.C. in Fig.~\ref{fig:ESI}) may be added to path b, and result in $\phi_0$ (constant time lag correspondingly) phase adjustment in $b$. Here we assume that the time lag from the BS splitter are the same for $D_1$ and $D_2$ by symmetric setup of the two detectors with respect to the BS. For the same reason, $\delta \tau$ are the same with respect to the two detectors, except for the sign. $\tau_p (t-\tau_0)$ and $\tau_p' (t-\tau_0)$ are the time delays caused by plasma. $\tau_p$ is time-dependent. $\tau_0$ is the free space propagation time between the plasma and the BS.

For $f_1(-) = f_2(+) = 1$, assume that $F_\nu$ is a Gaussian with a bandwidth $\Delta \omega$, $G^{(2)}(t_1;t_2)$ function is
\begin{equation}\begin{split}
G^{(2)} =&  T^2 |g[t_2 - t_1 -\tau_p(t_2) + \tau_p'(t_1)] |^2\\
&+R^2 |g[t_1 -t_2 +2 \delta \tau -\tau_p(t_1 +\delta \tau)+\tau_p'(t_2 -\delta \tau) ]|^2 \\
&-RT e^{-i \omega_p/2 (\tau_p'(t_2) -\tau_p'(t_1) +\tau_p(t_1) -\tau_p (t_2))} \\
& \times g[t_2 - t_1- \tau_p(t_2) + \tau_p'(t_1) ] \times \\
& g^*[t_1 -t_2 +2 \delta \tau -\tau_p(t_1 +\delta \tau)+\tau_p'(t_2 -\delta \tau)] +c.c.
\end{split}
\end{equation}
For a plasma with a linear growth (LG) in time delays,
\begin{equation}
\tau_p (t) =(a_0 + \frac{b}{2}) t + \tau_p, \tau_p'(t) = (a_0-\frac{b}{2}) t + \tau_p', 
\end{equation}
with $a_0$, $b$, $\tau_p$ and $\tau_p'$ being constants. Let $t_1 + t_2 = 2t$,  $t_2-t_1 = \tau$, $\tau_p-\tau_p' = \Delta \tau_p$, then
\begin{equation}\begin{split}
& G^{(2)} (LG) = \\
&  T^2 |g[(1- a_0)\tau - bt -\Delta \tau_p)] |^2\\
&+R^2 |g[-(1-a_0)\tau -bt +(2-b) \delta\tau -\Delta \tau_p ]|^2 \\
&-RT e^{i(\omega_p/2)b\tau}  g[(1-a_0)\tau - b t - \Delta \tau_p ]  \\
& \times g^*[-(1-a_0)\tau  - bt +(2-b) \delta \tau -\Delta \tau_p] \\
&+c.c.
\end{split}
\end{equation}

The mean joint detection probability or the average count rate for a detection window centered at $T_0$ is
\begin{equation}
R_c = \frac{1}{S_0} \int_{-\frac{S_0}{2}}^{\frac{S_0}{2}} dt_1 \int_{-\frac{S_0}{2}}^{\frac{S_0}{2}} dt_2 G^{(2)} S(t_1-t_2, \Delta S_{c})
\end{equation}
where $S(t_1-t_2, \Delta S_{c})$ is a concidence detection function, $\Delta S_{c}$ is the coincidence time window, the detection signal accumulation is over $S_0$.
We obtain,
\begin{equation}
\begin{split}
R_c (LG) = & C [1 - \frac{2RT}{R^2+T^2} \\
&\times B \cos \frac{\omega_p}{2}\frac{(b-b^2/2) \delta \tau }{1-a_0} \\
& \times e^{-\Delta \omega^2 [(1-b/2)\delta \tau - bt - \Delta \tau_p]^2 } ]
\end{split}
\end{equation}
with $B \propto e^{- \frac{\pi \omega^2_b b^2}{4 \Delta \omega^2 (1-a_0)^2}} $. For a steady plasma, $a_0 = b = 0$,
\begin{equation}
R_c = C \left[1 - \frac{2RT}{R^2+T^2} e^{-\Delta \omega^2 (\delta \tau - \Delta \tau_p)^2} \right].
\label{eq:rc}
\end{equation}

\begin{figure}[htbp] 
   \centering
   \includegraphics[width=3in]{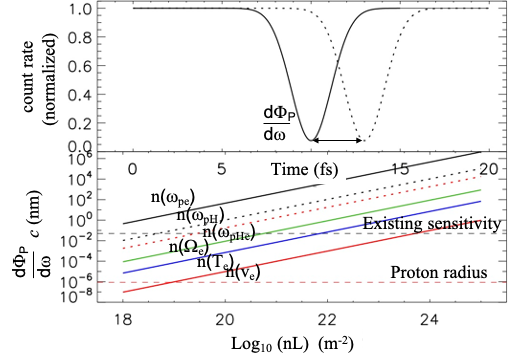} 
   \caption{(Top) Theoretical prediction of the plasma-induced shift in entangled photon correlation according to Eq.~(\ref{eq:rc}), The exact amount depends on plasma parameters as shown at the Bottom frame for different plasma effects as a function of line-integrated plasma density. The photon wavelength is assumed to be 1 $\mu$m.}
   \label{fig:SADip}
\end{figure}

The physical implication of Eq.~(\ref{eq:rc}) is consistent with the expectation that, changes in plasma refractive index will induce the shifts in position of the inverse Gaussian function, with the shift amount $\Delta \tau_p$. The normalized $R_c$ is shown in Fig.~\ref{fig:SADip}, for $R=0.45$ and $\Delta \omega = 2\pi/\tau_0$ with $\tau_0$ = 10 fs. The center of the dip position may be adjusted by inserting a phase object ($\Phi_0 = \omega \tau_b$) into path b in Fig.~\ref{fig:ESI}. The expected $\Delta \tau_p$ as a function of $\int n dl = nL$ for effects by different ions (H, He -- 3\% of electron density), magnetic field ($\Omega_e$), electron temperature, and electron flow are also shown in Fig.~\ref{fig:ESI}. Experimentally demonstrated sensitivity is used as the existing reference for expected sensitivity.  Further optimization schemes such as multi-path design, longer photon wavelengths (preferred for smaller $nL$), and higher entangled photon fluxes will increase sensitivities towards the equivalent length ($\Delta \tau_p c$) comparable to the proton radius. 

For the time-dependent cases $a_0, b \neq 0$. In this case, we expect both the time-dependent shift of the inverse Gaussian, slope = b/(1-b/2), and the broadening factor $\delta \omega (1 - b/2)$. We expect different sensitivities by choosing several setup configurations as summarized in Table.~\ref{bire}.
\begin{table} [thb!]
\centering
\caption{Relative phase shifts due to birefringence of a magnetized plasma for different setups. The two path-configurations are: [{\bf S}] for one of the two paths passing through the plasma, while the other path passing through a non-dispersive medium; [{\bf D}] for both paths through the plasma.}
\begin{tabular}{p{1.5cm}p{2.5cm}p{3.5cm} } 
\hline
Path  & symmetry & relative phase shift  \\
\hline
S & N & $n_R$  or  $n_L$   \\
D & Y & $n_R$, $n_R$; or $n_L$, $n_L$   \\
D & Anti-symmtry & $n_R$, $n_L$ \\
\hline
\end{tabular}
\label{bire}
\end{table}

In summary, through sub-picosecond timing by entangled photons and nonlocal intensity correlation, we introduce a quantum interferometer concept to measure small amplitude signals including fluctuations in turbulent and other plasmas. Analysis of a warm plasma dispersion function has revealed rich physics that is typically neglected and hard to measure presently because of $\omega^{-4}$-dependence on the photon frequency $\omega$.  Quantum metrology based on entangled photons and other non-classical source of light could open up new avenues to experiments on fundamental and complex problems such as anomalous transport in turbulent plasmas and predictive plasma science. This work is supported in part by the LANL LDRD program.


\begin{thebibliography}{99}
\bibitem{AAA:2013} J. Aasi {\it et al}, {\it Nat. Photonics} {\bf 7} (2013) 613; B. P. Abbott et al. {\it Phys. Rev. Lett.} {\bf 116}, 061102 (2016).
\bibitem{Ped:1999} M. A. Pedrosa, C. Hidalgo, B. A. Carreras, {\it et al}, {\it Phys. Rev. Lett.} {\bf 82}, 3621 (1999).
\bibitem{BDT:2003}  W. X. Ding, D. L. Brower, S. D. Terry, {\it et al}, {\it Phys. Rev. Lett.} {\bf 90} (2003) 035002.
\bibitem{MM:2011} J. E. Maggs and G. J. Morales, {\it Phys. Rev. Lett.} {\bf 107}, 185003 (2011).

\bibitem{HP:2005} G. J. M. Hagelaar and L. C. Pitchford,  {\it Plasma Sources Sci. Technol.} {\bf 14} (2005) 722-733.
\bibitem{Filbet:2006} F. Filbet {\it et al.} {\it SIAM J. Sci. Comput.} {\bf 28} (2006) 1029.

\bibitem{Evans:1976} D. E. Evans,  {\it Physica} {\bf 82C} (1976) 27-42.
\bibitem{Kunze:2005} H.-J. Kunze, {\it Plasma Diagnostics} in Lect. Notes. Phys. {\bf 670} (2005) 351-373.
\bibitem{DCB:2007} A. J. H. Donne, A. E. Costley, {\it et al}, `Chapter 7: Diagnostics,' {\it Nucl. Fusion} {\bf 47} (2007) S337-384.
\bibitem{DKK:2014} B. H. Deng, J. S. Kinley, K. Knapp {\it et al}, {\it Rev. Sci. Instrum.} {\bf 85} (2014) 11D401.

\bibitem{MW:1995} Mandel, L., and E. Wolf, 1995, Optical Coherence and Quantum Optics (Cambridge University Press, Cambridge).
\bibitem{SZ:1997} M. O. Scully, and M. S. Zubairy, 1997, {\it Quantum Optics} (Cambridge University Press, Cambridge, England).
\bibitem{Glauber:2007} R. Glauber, {\it Quantum  Theory  of  Optical  Coherence:  Selected Papers and Lectures}, (Wiley-VCH, Weinheim, 2007).
\bibitem{Shih:2011} Y. H. Shih, {\it An Introduction to Quantum Optics} (CRC Press, 2011).
\bibitem{Hut:2002} I. H. Hutchinson, {\it Principles  of Plasma  Diagnostics}, 2nd Ed. (Cambridge University Press, 2002).
\bibitem{HanB:1974} R. Hanbury Brown, `The intensity interferometer.' (Taylor \& Francis, 1974).

\bibitem{Shih:2003} Y. Shih, {\it Rep. Prog. Phys.} {\bf 66} (2003) 1009.
\bibitem{PCL:2012} J.-W. Pan {\it et al},  {\it Rev. Mod. Phys.} {\bf 84} (2012) 777.

\bibitem{KP:1966} C. F. Kennel and H. E. Petschek, {\it J. Geophys. Res.} {\bf 71}(1) (1966) 1.
\bibitem{Sti:1992} T. H. Stix, {\it Waves in Plasmas}, (AIP Press, 1992).

\bibitem{LKB:2018} A. Lyons, G. C. Knee, E. Bolduc et al,  {\it Sci. Adv.} {\bf 4} (2018) eaap9416.

\bibitem{HOM:1987} C. K. Hong, Z. Y. Ou, L. Mandel, {\it Phys. Rev. Lett.} {\bf 59}(18) (1987) 2044.
\bibitem{SA:1988} Y. H. Shih and C. O. Alley, {\it Phys. Rev. Lett.} {\bf 61} (1988) 2921.
\bibitem{SSR:1994b} Y. H. Shih, A. V. Sergienko, M. H. Rubin, T. E. Kiess, and C. O. Alley, {\it Phys. Rev. A} {\bf 50} (1994) 23.
\bibitem{DJM:1999} Dauler E., Jaeger G., Muller A., Migdall A. and Sergienko A. V., {\it J. Res. NIST} {\bf 104} (1999) 1.

\bibitem{GLM:2011} V. Giovannetti, S. Lloyd and L. Maccone, {\it Phys. Rev. Lett.} {\bf 96} 010401 (2006).
\bibitem{CTD:1980} C. M. Caves, K. S. Thorne, R. W. P. Drever, V. D. Sandberg, and M. Zimmermann, {\it Rev. Mod. Phys.} {\bf 52} (1980) 341; C. M. Caves, {\it Phys. Rev. D} {\bf 23} (1981) 1693.
\bibitem{HB:1993} M.    J.    Holland    and    K.    Burnett,  {\it Phys.   Rev.   Lett.} {\bf 71} (1993) 1355-1358.





\bibitem{RKS:1994} M. H. Rubin, D. N. Klyshko, Y. H. Shih and A. V. Sergienko, {\it Phys. Rev. A} {\bf 50} (1994) 5122.














\end{thebibliography}
\end{document}